\documentclass[11pt]{iopart}
\usepackage[latin1]{inputenc}
\usepackage[T1]{fontenc}
\usepackage{graphicx}
\usepackage{amsfonts}
\usepackage{hyperref}
\usepackage{color}
\usepackage{soul}
\usepackage{mathrsfs}  
\usepackage{cleveref}
\usepackage{etoolbox}
\crefname{equation}{}{}
\patchcmd{\numparts}{\addtocounter{equation}{1}}{\refstepcounter{equation}}{}{}

\begin{document}
	
	\title{Inferring comparative advantage via entropy maximization}
	
	%%% referees: 
	% Paolo Barucca
	% Fabio Caccioli
	% Riccardo Di Clemente
	% Stefano Guarino
	% Fabrizio Lillo
	% Giacomo Livan
	% Piero Mazzarisi
	% Davide Pirino
	
	\author{Matteo Bruno$^{1,2}$, Dario Mazzilli$^{2}$, Aurelio Patelli$^{2}$, Tiziano Squartini$^{3,4}$ and Fabio Saracco$^{2,3}$}
	\address{$^1$ SONY Computer Science Laboratories, Rome, Via Panisperna 89A, 00184 Rome, Italy}
	\address{$^2$ Enrico Fermi Research Center, via Panisperna 89a, 00184, Rome (Italy)}
	\address{$^3$IMT School for Advanced Studies, Piazza San Francesco 19, 55100 - Lucca (Italy)}
	\address{$^4$ Institute for Advanced Study (IAS), University of Amsterdam, Oude Turfmarkt 145, 1012 GC - Amsterdam (The Netherlands)}
	
	\date{\today}
	
	\begin{abstract}
		We revise the procedure proposed by Balassa to infer comparative advantage, which is a standard tool, in Economics, to analyze specialization (of countries, regions, etc.).
		Balassa's approach compares the export of a product for each country with what would be expected from a benchmark based on the total volumes of countries and products flows.
		Based on results in the literature, we show that the implementation of Balassa's idea generates a bias: the prescription of the maximum likelihood used to calculate the parameters of the benchmark model conflicts with the model's definition. 
		Moreover, Balassa's approach does not implement any statistical validation. Hence, we propose an alternative procedure to overcome such a limitation, based upon the framework of entropy maximisation and implementing a proper test of hypothesis: the `key products' of a country are, now, the ones whose production is significantly larger than expected, under a null-model constraining the same amount of information employed by Balassa's approach.
		What we found is that countries diversification is always observed, regardless of the strictness of the validation procedure. Besides, the ranking of countries' fitness is only partially affected by the details of the validation scheme employed for the analysis while large differences are found to affect the rankings of products Complexities.
		The routine for implementing the entropy-based filtering procedures employed here is freely available through the official Python Package Index \texttt{PyPI}.
	\end{abstract}
	
	\maketitle
	
	\section{Introduction}
	
	%%% intro
	The field of Economic Complexity has been steadily growing over the years. The first definition of an indicator for quantifying the complexity of the economy of a country is the so-called `Economic Complexity Index'~\cite{Hidalgo2009}; this first attempt has been followed by others, which have led to the definition of more sophisticated indicators such as the ones named Fitness~\cite{Tacchella2012,Cristelli2013}. Their goal is to measure the capabilities of a national industrial system. In Economics, the standard approach to this task is that of collecting all the relevant information about the economic system under analysis, such as data about labour, the presence of infrastructures, pollution, schooling, taxes and so on - a way of proceeding which is known to be affected by several problems~\cite{Cristelli2015}, as the one concerning the best way of combining such a large number of heterogeneous dimensions. 
	The Economic Complexity approach, instead, rests upon the assumption that country capabilities are proxied by their competitiveness and economic productive advantage.\\

	In standard Economics, the \emph{comparative advantage} of a nation accounts for the endogenous nature of labour (and capital): as a consequence, each country is incentivised to employ its labour force in the most remunerative sectors (`decided' by its endogenous capabilities) while trading goods belonging to the others. Within this setting, each nation will limit its production of commodities to the aforementioned sectors, hence `specialising' its export on few goods instead of `diversifying' its productive portfolio.
	Balassa realised that a country comparative advantage can be estimated through data analysis, in particular the one about trade exchanges. According to his intuition, a country has a comparative advantage in producing a given set of goods if it exports them `better than its competitors' whence the idea of estimating the competitiveness of a country, within a certain productive sector, by comparing its sector-specific export share with the one of the other world countries~\cite{Balassa1965,Balassa1967}. The index introduced by Balassa, named \emph{Revealed Comparative Advantage} (RCA) and extensively employed in the field of Economic Complexity~\cite{Hidalgo2009,Tacchella2018,Cristelli2013,Cristelli2015} does not follow from any formal derivation, being simply deduced from a quite general rationale. In fact, it can be shown that Balassa's RCA is a comparison between the empirical value of a country-specific product export and its expected counterpart under a null-model preserving the total export of each country and the total export of each product~\cite{Kunimoto1977,Vollrath1991,krantz2018}.
	It is worth mentioning that the application of RCA goes far beyond international trade nowadays. Sometimes with different names, it has been applied to patenting data for technologies~\cite{sbardella2018green,de2022trickle}, employment data for sectoral studies~\cite{sbardella2017economic} and scientific publication for science production \cite{patelli2023geography,pugliese2019unfolding}.\\ 
	
	In the present paper, we revise the Balassa's procedure. Firstly, we show that it is a comparison between the empirical value of a country-specific product export and the approximation of the outcome of the entropy-based Bipartite Weighted Configuration Model (\emph{BiWCM}, introduced in~\cite{digangi2018}) in the so-called `sparse' regime; the ingredients of the Bipartite Weighted Configuration Model are the ones originally used by Balassa to define RCA.\\ 
	Secondly, we will show that RCA is intrinsically biased. It was proved in the literature that there exists a set of null-models in which the definition of the null-model does not meet the maximum likelihood condition and that are, therefore,  biased~\cite{Garlaschelli2008}. Remarkably, RCA falls in this set, while ``standard" entropy-based null-models (as BiWCM) are safe. Notably, when calculating the fitnesses of countries, i.e. the ability of countries of exporting the most exclusive products, the error made by using RCA is limited: the rankings obtained using different methods are highly correlated. Instead, the situation is different when considering the products' complexities: in this case, the correlations are much more limited.\\ 
	Third, we overcome the intrinsic limitation of Balassa's recipe, which prescribes to compare an empirical value with its expected counterpart (as noticed in~\cite{krantz2018}, an overall poor validation from a statistical point of view), by proposing a proper test of hypothesis to be implemented by calculating the p-value of a country-specific product export. Specifically, the third point will be addressed by using probability distributions derived by maximising Shannon entropy in a constrained fashion, with the constraints representing the empirical properties to be preserved while randomising everything else~\cite{Cimini2018}: in the specific case, the constraints coincide with the ingredients employed by Balassa to define the RCA, i.e. the total export per country and per product.\\
	
	The sections of this paper are organized as follows: `Methods' section is devoted to introduce maximum entropy benchmarks (in particular, two variants of the Bipartite Configuration Model for weighted networks) and re-framing the RCA-based validation procedure within such a framework. The `Results' section is devoted to investigate the differences between the two different validation procedures (i.e. either the comparison of the empirical value with the statistical model's mean or the statistical validation) induced by the models considered for the present analysis. Finally, the `Conclusions' section is devoted to summarise and discuss our results as well as some of their implications.

	\section{Methods}
	
	In the present section, we discuss entropy-based null-models for the analysis of complex networks and apply such a framework to tackle the problem of providing a statistically sound validation of a country's production. Specifically, we will consider the Bipartite Weighted Configuration Model with both discrete and continuous weights, respectively with the names BiWCM$_d$ and BiWCM$_c$.
	Let us finally remark that BiWCM$_d$ was introduced earlier in the literature~\cite{digangi2018}.
	
	\subsection{Entropy-based null-models: general framework}\label{ssec:genframe}
	
	Entropy-based null-models represent benchmarks that are maximally random except for the constraints defining them. Let us start by defining the ensemble of graphs $\mathcal{G}$, including all graphs having the same number of nodes of the real system (in the case of bipartite networks, the same number of nodes per layer) and a number of links that varies from zero to the maximum. Then, let us `equip' this ensemble with a probability $P(G)$, where $G\in\mathcal{G}$ is a member of the ensemble, and whose associated Shannon entropy reads
	
	\begin{equation}
		S=-\sum_{G\in\mathcal{G}}P(G)\ln P(G).
	\end{equation}
	
	If $\vec{C}(G)$ is a vector of quantities associated with each graph, we can maximise Shannon entropy while constraining the average values of $\vec{C}$ over the ensemble itself. The method of Lagrange multipliers amounts to maximising $S'$ defined as 
	
	\begin{equation}\label{eq:Sp_genframe}
		S'=S+\vec{\theta}\cdot(\vec{C}^\bullet-\langle\vec{C}\rangle)+\gamma\left(1-\sum_{G\in\mathcal{G}}P(G)\right)
	\end{equation}
	where $\vec{\theta}$ is the vector of Lagrange multipliers associated with the constraints represented by $\vec{C}$, $\vec{C}^\bullet$ is the value of the constraint and $\gamma$ is the Lagrange multiplier associated with the normalization of the probability~\cite{Park2004}. The constrained maximization returns
	
	\begin{equation}\label{eq:ERG_p}
		P(G)=\frac{e^{-\vec{\theta}\cdot\vec{C}(G)}}{Z(\vec{\theta})}
	\end{equation}
	where $Z=e^{\gamma+1}=\sum_{G\in\mathcal{G}}e^{-\vec{\theta}\cdot\vec{C}(G)}$ is the partition function.

	So far, we do not know the numerical value of the Lagrangian multipliers, since Eq.(\ref{eq:ERG_p}) describes only the exponential functional form of $P(G)$.
	If we want to tailor our null-model on the real system --for instance, to spot behaviours that deviate from what is encoded in the constraints--, we require that $\langle\vec{C}\rangle$ equate the empirical values of the constraints $\vec{C}^*$, or, otherwise stated to put $\vec{C}^\bullet=\vec{C}^*$ in Eq.(\ref{eq:Sp_genframe}) (hereafter, quantities denoted by an asterisk $*$ will indicate values measured on the real system). Remarkably, it can be proven that, in this case, the maximum-likelihood estimation of the Lagrange multipliers leads precisely to the condition~\cite{Garlaschelli2008}:
	
	\begin{equation}\label{eq:maxlikelihood}
		\langle \vec{C}\rangle=\vec{C}^*.
	\end{equation}
	Solving Eq.(\ref{eq:maxlikelihood}) typically implies solving a system of non-linear, coupled equations whose number is proportional to the number of nodes of the network under analysis: in what follows, we will describe an algorithm to numerically solve Eq.(\ref{eq:maxlikelihood}).
	
	\subsection{Entropy-based null-models for bipartite weighted networks}\label{ssec:bi*cm}
	
	Let us call the two layers of a bipartite network $\top$ and $\bot$ and indicate their dimensions as $N_\top$ and $N_\bot$, respectively. A bipartite weighted network is completely defined by its weighted biadjacency matrix $\mathbf{W}$, i.e. a $N_\top\times N_\bot$ matrix whose generic entry $w_{i\alpha}$ is 0 if no connection is present between $i\in\top$ and $\alpha\in\bot$; otherwise, $w_{i\alpha}$ is positive and captures the intensity of the interaction between the same two nodes. In case of discrete weights, $w_{i\alpha}$ is an integer and coincides with the number of parallel edges between $i$ and $\alpha$; otherwise, $w_{i\alpha}$ is a real number. The strengths are defined as the marginals of $\mathbf{W}$, i.e. as $s_i=\sum_{\alpha\in\bot}w_{i\alpha}$, $\forall\:i$ and $\sigma_\alpha=\sum_{i\in\top}w_{i\alpha}$, $\forall\:\alpha$.
	
	\subsubsection{Discrete Bipartite Weighted Configuration Model}
	
	The Discrete Bipartite Weighted Configuration Model was introduced in~\cite{digangi2018}. Following the procedure sketched in Subsection~\ref{ssec:genframe}, we impose the node strengths of both layers as constraints:
	
	\begin{equation}\label{eq:const_biwcms}
		\vec{\theta}\cdot\vec{C}(G_W)=\sum_i\theta_i s_i(G_W)+\sum_\alpha \eta_\alpha\sigma_\alpha(G_W)=\sum_{i,\alpha}(\theta_i+\eta_\alpha)w_{i\alpha}(G_W)
	\end{equation}
	where $G_W$ is a generic weighted graph belonging to the ensemble $\mathcal{G}_W$, $\theta_i$ and $\eta_\alpha$ are the Lagrange multipliers associated, respectively, with $s_i$ and $\sigma_\alpha$. Equation (\ref{eq:ERG_p}), then, reads
	
	\begin{equation}
		P_{\textnormal{BiWCM}_d}(G_W)=\frac{\prod_{i,\alpha} e^{-(\theta_i+\eta_\alpha)w_{i\alpha}(G_W)}}{Z_{\textnormal{BiWCM}_d}};
	\end{equation}
	since weights are discrete, one finds the expression
	
	\begin{eqnarray}
		\label{eq:Z_bimcm}
		Z_{\textnormal{BiWCM}_d}&=\sum_{G_W\in\mathcal{G}_W}\prod_{i,\alpha} e^{-(\theta_i+\eta_\alpha)w_{i\alpha}(G_W)}=\prod_{i,\alpha}\sum_{G_W\in\mathcal{G}_W} e^{-(\theta_i+\eta_\alpha)w_{i\alpha}(G_W)}\nonumber\\
		&=\prod_{i,\alpha}\sum_{w_{i\alpha}=0}^\infty e^{-(\theta_i+\eta_\alpha)w_{i\alpha}}=\prod_{i,\alpha}\frac{1}{1-e^{-(\theta_i+\eta_\alpha)}}
	\end{eqnarray}
	in turn inducing
	
	\begin{equation}\label{eq:p_bimcm}
		P_{\textnormal{BiWCM}_d}(G_W)=\prod_{i,\alpha}e^{-(\theta_i+\eta_\alpha)w_{i\alpha}(G_W)}[1-e^{-(\theta_i+\eta_\alpha)}]
	\end{equation}
	which allows us to interpret the probability of a graph $G_W\in\mathcal{G}_W$ as the product of probabilities whose generic functional form reads $q_{\textnormal{BiWCM}_c}(w_{i\alpha})=e^{-(\theta_i+\eta_\alpha)w_{i\alpha}}[1-e^{-(\theta_i+\eta_\alpha)}]$.
	
	In order to get the numerical values of the Lagrange multipliers $\{\theta_i\}_{i\in\top}$ and $\{\eta_\alpha\}_{\alpha\in\bot}$, we impose $\langle s_i\rangle_{\textnormal{BiWCM}_d}=\sum_\alpha\langle w_{i\alpha}\rangle_{\textnormal{BiWCM}_d}=s_i^*$, $\forall\:i$ and that $\langle \sigma_\alpha\rangle_{\textnormal{BiWCM}_d}=\sum_i\langle w_{i\alpha}\rangle_{\textnormal{BiWCM}_d}=\sigma_\alpha^*$, $\forall\:\alpha$\footnote{As in subsection~\ref{ssec:genframe}, quantities denoted by an asterisk $*$ are the ones observed in the real system.}, i.e.
	
	\begin{equation}\label{eq:likelihood_bimcm}
		\left\{
		\begin{array}{cc}
			s_i^*&=\sum_\alpha\frac{e^{-(\theta_i+\eta_\alpha)}}{1-e^{-(\theta_i+\eta_\alpha)}},\quad\forall\:i\\
			\sigma_\alpha^*&=\sum_i\frac{e^{-(\theta_i+\eta_\alpha)}}{1-e^{-(\theta_i+\eta_\alpha)}},\quad\forall\:\alpha
		\end{array}
		\right..
	\end{equation}
	As stressed in the subsection above, the condition for the numerical determination of the Lagrange multipliers is equivalent at maximising the likelihood~\cite{Garlaschelli2008} (please refer to Appendix~\ref{app:BiMCM} for the algorithms used to numerically solve the system above). Since $q_{\textnormal{BiWCM}_d}(w_{i\alpha})$ obeys a geometric distribution, the statistical validation of a generic, empirical entry $w_{i\alpha}^*$ leads to recover an expression for the associated p-value reading
	
	\begin{eqnarray}\label{eq:pval_bimcm}
		\textnormal{p-value}_{\textnormal{BiWCM}_d}(w_{i\alpha}^*)&=\sum_{w\geq w_{i\alpha}^*}^{+\infty}q_{\textnormal{BiWCM}_d}(w) \\
		&=\sum_{w\geq w_{i\alpha}^*}^{+\infty}e^{-(\theta_i+\eta_\alpha)w}[1-e^{-(\theta_i+\eta_\alpha)}]=e^{-(\theta_i+\eta_\alpha)w_{i\alpha}^*}.\nonumber
	\end{eqnarray}
	
	\subsubsection{Continuous Bipartite Weighted Configuration Model}
	
	Let us now consider our entropy-based framework in case weights are treated as continuous variables. To the best of our knowledge, such a null-model has not been defined yet and represents the bipartite analogue of the model introduced in~\cite{Parisi2020}. The Hamiltonian defined in Eq.(\ref{eq:const_biwcms}) remains the same, the only difference being the nature of the weights. In order to calculate the partition function, we need to integrate, now, finding
	
	\begin{eqnarray}\label{eq:Z_biwcm}
		Z_{\textnormal{BiWCM}_c}&=\sum_{G_W\in\mathcal{G}_W} \prod_{i,\alpha}e^{-(\theta_i+\eta_\alpha)w_{i\alpha}(G_W)}=\prod_{i,\alpha}\int_0^\infty e^{-(\theta_i+\eta_\alpha)w_{i\alpha}}dw_{i\alpha}\nonumber\\
		&=\prod_{i,\alpha}\frac{1}{\theta_i+\eta_\alpha}
	\end{eqnarray}
	an expression inducing a probability per graph reading
	
	\begin{equation}\label{eq:p_biwcm}
		P_{\textnormal{BiWCM}_c}(G_W)=\prod_{i,\alpha}e^{-(\theta_i+\eta_\alpha)w_{i\alpha}(G_W)}[\theta_i+\eta_\alpha];
	\end{equation}
	therefore, the probability density function associated with the entry $w_{i\alpha}$ is simply $q_{\textnormal{BiWCM}_c}(w_{i\alpha})=e^{-(\theta_i+\eta_\alpha)w_{i\alpha}}[\theta_i+\eta_\alpha]$.
	
	Again, to obtain the values of the Lagrange multipliers $\{\theta_i\}_{i\in\top}$ and $\{\eta_\alpha\}_{\alpha\in\bot}$, we impose that
	
	\begin{equation}\label{eq:likelihood_biwcm}
		\left\{
		\begin{array}{cc}
			s_i^*&=\sum_\alpha\frac{1}{\theta_i+\eta_\alpha},\quad\forall\:i\\
			\sigma_\alpha^*&=\sum_i\frac{1}{\theta_i+\eta_\alpha},\quad\forall\:\alpha
		\end{array}
		\right.
	\end{equation}
	and such a condition is equivalent to maximising the likelihood~\cite{Garlaschelli2008} (please refer to Appendix~\ref{app:BiWCM} for the algorithms used to solve the system above numerically). Since $q_{\textnormal{BiWCM}_c}(w_{i\alpha})$ obeys an exponential distribution, the statistical validation of a generic, empirical entry $w_{i\alpha}^*$ leads to recover an expression for the associated p-value reading
	
	\begin{eqnarray}\label{eq:pval_biwcm}
		\textnormal{p-value}_{\textnormal{BiWCM}_d}(w_{i\alpha}^*)&=\int_{w\geq w_{i\alpha}^*}^{+\infty}q_{\textnormal{BiWCM}_c}(w)\, dw\\
		&=(\theta_i+\eta_\alpha)\int_{w\geq w_{i\alpha}^*}^{+\infty}e^{-(\theta_i+\eta_\alpha)w}\, dw=e^{-(\theta_i+\eta_\alpha)w_{i\alpha}^*}.\nonumber
	\end{eqnarray}
	Even if the functional form of Eq.(\ref{eq:pval_biwcm}) is the same of Eq.(\ref{eq:pval_bimcm}), it is worth remembering that, in general, the numerical values of the Lagrange multipliers are different in the two cases.
	
	\subsubsection{Final remarks}\label{sssec:equiv_bimcm_biwcm}
	First, we have implemented the BiWCM$_d$ and the BiWCM$_c$ solvers distributed as updates of the {\tt NEMtropy}\footnote{\url{https://github.com/nicoloval/NEMtropy}} and the {\tt bicm}\footnote{\url{https://github.com/mat701/BiCM}} Python packages.
	
	Second, in the `dense' regime, i.e. for `small' Lagrange multipliers, the BiWCM$_d$ and the BiWCM$_c$ are indistinguishable: in fact, in the `dense' regime, Eq.(\ref{eq:p_bimcm}) can be approximated as 
	
	\begin{eqnarray}
		P_{\textnormal{BiWCM}_d}(G_W)&=\prod_{i,\alpha}e^{-(\theta_i+\eta_\alpha)w_{i\alpha}(G_W)}[1-e^{-(\theta_i+\eta_\alpha)}] \\
		&\simeq\prod_{i,\alpha}e^{-(\theta_i+\eta_\alpha)w_{i\alpha}(G_W)}[\theta_i+\eta_\alpha], \nonumber
	\end{eqnarray}
	i.e. an expression bearing the same functional form of Eq.(\ref{eq:p_biwcm}). As a consequence, Eq.(\ref{eq:likelihood_bimcm}) becomes
	
	\begin{eqnarray}
		s_i^*=\langle s_i\rangle_{\textnormal{BiWCM}_d}&=\sum_\alpha\frac{e^{-(\theta_i+\eta_\alpha)}}{1-e^{-(\theta_i+\eta_\alpha)}}\simeq\sum_\alpha\frac{1}{1-\big(1-(\theta_i+\eta_\alpha)\big)}\nonumber\\
		&=\sum_\alpha\frac{1}{\theta_i+\eta_\alpha},
	\end{eqnarray}
	which is solved by the Lagrange multipliers associated with the BiWCM$_c$ (see Eq.(\ref{eq:likelihood_biwcm})). An analogous consideration holds for $\sigma_\alpha$. In the systems we tested, the numerical differences between the Lagrange multipliers of the BiWCM$_d$ and the ones of the BiWCM$_c$ are of the order of $10^{-8}$, implying that the difference between their probabilities is out of the numerical precision, using double precision floats: therefore, in what follows we will just focus on the performance of the BiWCM$_c$.
	
	Second, we explicitly notice that there exist entropy-based null-models that are able to account both for the weights and for the topology (i.e. for the number of connections of each node, disregarding its strength) of the network: see, for instance~\cite{Mastrandrea2014}. Nevertheless, we prefer to neglect this information, here, as it was missing from Balassa's original proposal as well.
	
	\subsection{Revising the RCA-based validation}
	
	\subsubsection{Balassa's RCA-based thresholding}
	
	As mentioned in the Introduction, Balassa's thresholding procedure prescribes to compare the generic entry $w_{i\alpha}^*$ with the threshold
	
	\begin{equation}\label{eq:rca}
		w^\textnormal{Balassa}_{i\alpha}=\frac{s_i^*\sigma_\alpha^*}{W^*},
	\end{equation}
	where $W^*=\sum_{i\in\top}s_i^*=\sum_{\alpha\in\bot}\sigma_\alpha^*$ and retain the corresponding link if the empirical value equals or exceeds it. In formulas
	
	\begin{equation}
		m_{i\alpha}=\left\{
		\begin{array}{ll}
			1 & \textnormal{RCA}_{i\alpha}=\frac{w_{i\alpha}^*}{w^\textnormal{Balassa}_{i\alpha}}\ge1\\
			0 & \textnormal{otherwise}
		\end{array}
		\right..
	\end{equation}
	Looking back at Eq.s(\ref{eq:likelihood_bimcm}), we realize that $w^\textnormal{Balassa}_{i\alpha}$ coincides with $\langle w_{i\alpha}\rangle_{\textnormal{BiWCM}_d}$ in the `sparse' regime, i.e. for `large' Lagrange multipliers, which ensure that $1-e^{-(\theta_i+\eta_\alpha)}\simeq1$: in this case, in fact, $\theta_i+\sigma_\alpha\simeq\ln W^*-\ln s_i^*-\ln \sigma_\alpha^*$ and
	
	\begin{equation}
		\langle w_{i\alpha}\rangle_{\textnormal{BiWCM}_d}\simeq\frac{s_i^*\sigma_\alpha^*}{W^*}=w^\textnormal{Balassa}_{i\alpha}.
	\end{equation}
	
	In this sense, Balassa's validation adopts the approximated BiWCM$_d$ as a benchmark, an identification that does not hold when the BiWCM$_c$ is considered (in fact, the probability distribution of the BiWCM$_d$ is indistinguishable from the one of the BiWCM$_c$ only in the `dense' regime, i.e. for `small' $\theta_i$ and $\eta_\alpha$).

	\subsubsection{`Dressed' RCA-based validation}
	
	Entropy-based null-models maximise Shannon entropy while constraining the expected value of a number of network properties: as such, they represent a quite natural framework to carry out unbiased inference. In the present section, we exploit their statistical properties by `dressing' Balassa's threshold with the probability distributions characterising the BiWCM$_d$ and the BiWCM$_c$. Operatively, we follow the strategy proposed in~\cite{digangi2018}\footnote{Here, the model induced by such a strategy is named Maximum-Entropy Capital Asset Pricing Model (MECAPM). The present paper, however, does not deal with financial systems, whence our choice of renaming it Discrete Maximum-Entropy RCA (MERCA$_d$). Analogously, its continuous counterpart will be named Weighted Maximum Entropy RCA (MERCA$_c$).} and solve the equations 
	
	\begin{equation}
		\langle w_{i\alpha}\rangle=w_{i\alpha}^\textnormal{Balassa},\quad\forall\:i,\alpha
	\end{equation}
	($\langle\cdot\rangle$ are the average over the ensemble with the probability defined by the present null-model) to obtain the numerical values of the parameters in both cases and implement the corresponding validation algorithm.\\
	
	\paragraph*{Discrete Maximum-Entropy RCA (MERCA$_d$) model.} Calculating the probability per graph within the MERCA$_d$ framework is pretty straightforward, following the same steps used to get the probabilities per graph in Eq.(\ref{eq:p_bimcm}). In fact, 
	
	\begin{eqnarray}
		P_{\textnormal{MERCA}_{d}}(G_W)&=\prod_{i,\alpha}\left[\frac{s_i^*\sigma_\alpha^*}{W^*+s_i^*\sigma_\alpha^*}\right]^{w_{i\alpha}(G_W)}\left[\frac{W^*}{W^*+s_i^*\sigma_\alpha^*}\right]\nonumber\\
		&=\prod_{i,\alpha}\left[\frac{w_{i\alpha}^\textnormal{Balassa}}{1+w_{i\alpha}^\textnormal{Balassa}}\right]^{w_{i\alpha}(G_W)}\left[\frac{1}{1+w_{i\alpha}^\textnormal{Balassa}}\right];
	\end{eqnarray}
	as in the case of the BiWCM$_d$, the weight-specific probability distribution is a geometric one. P-values can be calculated as in Eq.(\ref{eq:pval_bimcm}), i.e.
	
	\begin{equation}
		\textnormal{p-value}_{\textnormal{MERCA}_{d}}(w_{i\alpha}^*)=\left[\frac{s_i^*\sigma_\alpha^*}{W^*+s_i^*\sigma_\alpha^*}\right]^{w_{i\alpha}^*}.
	\end{equation}
	
	\paragraph*{Continuous Maximum-Entropy RCA (MERCA$_c$) model.} In this second case, the derivation of the probability per graph follows from Eq.(\ref{eq:p_biwcm}). In fact,
	
	\begin{eqnarray}
		P_{\textnormal{MERCA}_c}(G_W)&=\prod_{i,\alpha}e^{-\frac{W^*}{s_i^*\sigma_\alpha^*}w_{i\alpha}(G_W)}\left[\frac{W^*}{s_i^*\sigma_\alpha^*}\right]\nonumber\\
		&=\prod_{i,\alpha}e^{-\frac{w_{i\alpha}(G_W)}{w_{i\alpha}^\textnormal{Balassa}}}\left[\frac{1}{w_{i\alpha}^\textnormal{Balassa}}\right]
	\end{eqnarray}
	and the p-value associated with the generic weight reads:
	
	\begin{equation}
		\textnormal{p-value}_{\textnormal{MERCA}_c}(w_{i\alpha}^*)=e^{-\frac{W^*}{s_i^*\sigma_\alpha^*}w_{i\alpha}(G_W)}.
	\end{equation}

	\subsubsection{Final remarks}

	Let us comment on the maximum-likelihood procedure introduced in subsection~\ref{ssec:genframe}. As stressed in~\cite{Garlaschelli2008}, likelihood maximisation reveals the presence of biases whenever the condition imposed to maximise the likelihood does not agree with the requirement of the model definition.
	In these cases, in fact, the maximisation of the likelihood returns a different set of conditions to be satisfied by the model parameters. Instead, ``usual'' entropy-based null-models, i.e. those for which the constraints are fixed by the observed network (i.e. $\langle\vec{C}\rangle=\vec{C}^*$) are unbiased since likelihood maximisation can be proven to be equivalent to the aforementioned set of conditions. MERCA models, while following from the maximisation of the entropy, lie somewhere in between the two cases above, since likelihood maximisation would require $\langle w_{i\alpha}\rangle_{\textnormal{MERCA}_\circ}=w_{i\alpha}^*$, $\forall\:i,\alpha$ while the models impose $\langle w_{i\alpha}\rangle_{\textnormal{MERCA}_\circ}=s_i^*\sigma_\alpha^*/W^*$, $\forall\:i,\alpha$.
	In fact, if we are interested in imposing constraints on weights, using some hypothesis is necessary since it ensures that our validation procedure will not lead to trivial results, as it would happen by constraining the expected value of each pair-specific weight to match the empirical one. Otherwise stated, we are interested in finding significant `deviations' from a model's expectations while solving the maximum-likelihood set of conditions would not lead, by definition, to any significant deviation. In any case, MERCA models satisfy the constraints defining BiWCMs:
	
	\begin{equation}
		\langle s_i\rangle_{\textnormal{MERCA}_\circ}=\sum_\alpha\langle w_{i\alpha}\rangle_{\textnormal{MERCA}_\circ}=\sum_\alpha\frac{s_i^*\sigma_\alpha^*}{W^*}=s_i^*,\quad\forall\:i
	\end{equation}
	and analogously for $\langle \sigma_\alpha\rangle_{\textnormal{MERCA}_\circ}$, $\forall\:\alpha$. In this sense, the MERCA models contain the same constraints of BiWCM models, but they are, at the same time, more restrictive (since constraints are per link, instead of per node) and not ``centred" on the real system (since the likelihood maximisation condition does not match with the definition of the model).\\
	
	Second, as for the BiWCM$_d$ and the BiWCM$_c$, we observed no differences between the probabilities calculated using the MERCA$_d$ model and the MERCA$_c$ model, up to the numerical precision of a double-float representation. Therefore, in what follows we will just focus on the performance of the MERCA$_c$ model.
	
	\subsection{Testing multiple hypotheses via FDR}
	
	When validating several hypotheses, the effect due to the number of tests carried out at the same time plays a role and not accounting for it may lead to a large number of false positives~\cite{Benjamini1995}. Among the different techniques that have been proposed to correctly tackle this problem, we opt for the one named \emph{False Discovery Rate} (FDR), able to control the percentage of false positives~\cite{Benjamini1995}. In a nutshell, this procedure prescribes sorting the p-values associated with the hypotheses to be tested (say $n$) in increasing order, i.e.
	
	\begin{equation}
		\textnormal{p-value}_1\le\textnormal{p-value}_2\le\cdots\le\textnormal{p-value}_n.
	\end{equation}
	Upon choosing a significance level (say $\alpha$), a global threshold $\alpha_\textnormal{FDR}=\hat{i}\alpha/n$ remains defined where $\hat{i}$ is the greatest $i$ that satisfies the relationship
	
	\begin{equation}
		\textnormal{p-value}_i\leq\alpha_\textnormal{FDR}
	\end{equation}
	(to be noticed that $\alpha$ represents the significance level of the whole set of hypotheses, every single link being validated by employing the lower threshold $\alpha_\textnormal{FDR}$). In what follows, we will set $\alpha=0.05$.
	
	\subsection{Comparing different validation procedures}
	
	Balassa's original procedure can be generalised in different ways:
	
	\begin{enumerate}
		\item the first one prescribes to deem country $i$ as `competitive' in exporting product $\alpha$ if the amount of the exported product is simply above the average. In formulas,
		\begin{equation}
			m^{\mu-\textnormal{null-model}}_{i\alpha} = \left\{
			\begin{array}{cl}
				1 & \textnormal{if } w^*_{i\alpha} \geq \langle w_{i\alpha} \rangle_\textnormal{null-model}\\
				0 & \textnormal{otherwise}
			\end{array}
			\right.,
		\end{equation}
		where $\mu-\textnormal{null-model}$ indicates that we are using the mean of the distribution defining the specific `null-model';
		
		\item the second one prescribes to deem country $i$ as `competitive' in exporting product $\alpha$ if the amount of the exported product is statistically significant. In formulas,
		\begin{equation}
			m^{\alpha-\textnormal{null-model}}_{i\alpha} = \left\{
			\begin{array}{cl}
				1 & \textnormal{if } \textnormal{p-value}_\textnormal{null-model}(w^*_{i\alpha})\leq\alpha_\textnormal{FDR}\\
				0 & \textnormal{otherwise}
			\end{array}
			\right.,
		\end{equation}
		where $\alpha-\textnormal{null-model}$ indicates that we are carrying out a proper statistical validation, comparing the observations with the distribution defining the specific `null-model'.
	\end{enumerate}
	
	Summarising, we have 2 different validation procedures (the $\mu-$ and the $\alpha-$ones) and 2 different null-models (the MERCA$_c$ and the BiWCM$_c$, since in the regime of our analyses their discrete versions are indistinguishable from them).
	
	\section{Results}
	
	Our analysis focuses on the bipartite representation of International Trade, i.e. the bipartite network of countries (on  one layer) and exported products (on the opposite layer). Country $i$ is connected with product $\alpha$ if $i$ exports $\alpha$, the weight $w_{i\alpha}$ is the volume of the export in dollars. More concretely, we will consider the COMTRADE export \footnote{https://comtradeplus.un.org/}, classified according to HS2012 and pre-processed. The overall numbers of countries and categories of products are $169$ and $5,206$, respectively. Moreover, we will just focus on the matrices obtained for 2015 as the other years are basically characterised by the same patterns.
	
	\begin{figure}[t!]
		\includegraphics[width=\textwidth]{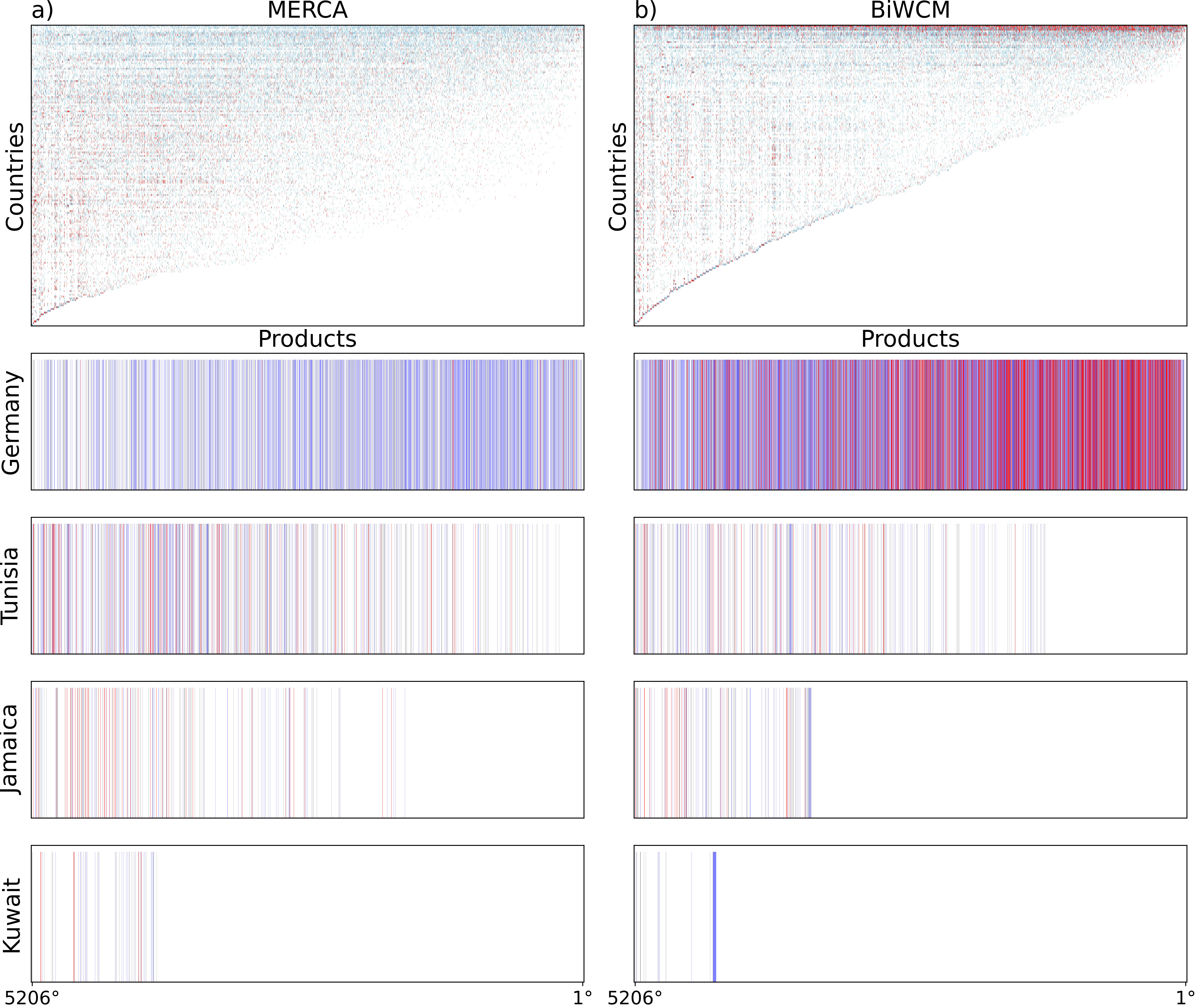}
		\caption{Top panels: Matrices representing the adjacency matrix of the Bipartite Network of the export of International Trade for the year 2015 of our dataset. Panel a) shows the $\mathbf{M}^{\mu-\textnormal{MERCA}_c}=\mathbf{M}^{\mu-\textnormal{RCA}}$ matrix: blue dots are validated exclusively by $\mu-\textnormal{MERCA}_c$, red ones are validated also by $\alpha-\textnormal{MERCA}_c$. Panel b) shows the $\mathbf{M}^{\mu-\textnormal{BiWCM}_c}$ matrix and the color codes the statistical validation: blue dots are validated exclusively by $\mu-$BiWCM$_c$, red ones are validated also by $\alpha-$BiWCM$_c$. Each matrix is re-ordered following the Fitness and Complexity ranking evaluated on the $\mu-$validated matrix. Bottom panels: bars representing the basket of products of a few samples from the matrices above.}%\\ 
		\label{fig:matrices}
	\end{figure}
	
	\subsection{Comparing validated matrices}
	
	First, let us notice that different validation methods induce different link densities: as Fig.~\ref{fig:matrices} shows, $\mu-$MERCA$_c$ (i.e. Balassa's RCA-based validation, panel a, blue dots) induces a connectance around 0.10 throughout the entire dataset, while $\mu-$BiWCM$_c$ (panel b, blue dots) induces a connectance which is steadily around 0.09. Being defined by a set of $N_\top\times N_\bot$ conditions on the weights, one may expect the $\mu-$MERCA$_c$ to provide a stricter validation than the BiWCM$_c$ which, in turn, is defined by a set of `just' $N_\top+N_\bot$ conditions. However, while the constraints defining the BiWCM$_c$ reflect properly empirical quantities, this is not the case for the MERCA$_c$: in fact, the $N_\top\times N_\bot$ conditions reading $\langle w_{i\alpha}\rangle_{\textnormal{MERCA}_c}=s_i^*\sigma_\alpha^*/W^*$, $\forall\:i,\alpha$ provide estimations which may perform poorly in reproducing the corresponding empirical quantities.
	
	The situation is different when coming to consider the $\alpha-$validation: indeed, the $\alpha-$MERCA$_c$ (panel a, red dots) validates $18,861$ links against the $19,938$ ones validated by the $\alpha-$BiWCM$_c$ (panel b, red dots). In principle, one would have expected to observe a larger number of links populating the matrix validated by employing the $\alpha-$MERCA$_c$.

	As the number of p-values which are lower than, say, $0.05$ is slightly larger for MERCA$_c$ ($38,833$) than for BiWCM$_c$ ($38,732$), such an effect is due to correcting for multiple hypotheses (it can be shown that p-values differing for less than $\alpha/n$ are either both validated or both non validated by FDR~\cite{Saracco2017}).
	
	For what concerns the number of empty columns, $\alpha-$MERCA$_c$ does not validate 633 products while $\alpha-$BiWCM$_c$ does not validate 65 products. However, non-validated columns (under both $\alpha-$MERCA$_c$ and $\alpha-$BiWCM$_c$) are distributed across the entire range of strengths, even if they are more present among the largest ones. 
	
	Overall, nearly two-thirds of the links validated by the $\mu-$MERCA$_c$ are also validated by the $\mu-$BiWCM$_c$ and roughly half of the links validated by the $\alpha-$MERCA$_c$ are also validated by the $\alpha-$BiWCM$_c$.\\
	
	The validated matrices in Fig.~\ref{fig:matrices} are reordered according to the fitness and complexity rankings (Appendix~\ref{app:fitness}), respectively for countries and products, calculated on the $\mu-$validated matrices.
	It is worth to notice that the triangular shape is emphasised more within the framework induced by $\mu-$BiWCM$_c$ than within the one induced by $\mu-$MERCA$_c$: in fact, the stable Nested Overlap and Decreasing Fill (sNODF)~\cite{Mariani2019}, i.e. a variation of the measure proposed in~\cite{AlmeidaNeto2008}, amounts at 0.24 when computed on the matrices validated by the $\mu-$BiWCM$_c$ and 0.21 when computed on the matrices validated by the $\mu-$MERCA$_c$. Such a result still holds true when a stricter validation procedure is employed: in fact, the $\alpha-$validated links in the network still display a clearly nested structure with the value of sNODF being larger for the BiWCM$_c$ (0.109) than for MERCA$_c$ (0.056).\\
	
	In Fig.~\ref{fig:matrices}, we have coloured the $\mu-$validated links in blue and those which are also $\alpha-$validated in red, each line representing a validated export whose thickness is proportional to the corresponding total amount (products are reordered according to the complexity ranking calculated on the corresponding matrix). As the export baskets of the single countries confirm, the validated matrix using the BiWCM$_c$ is sparser than the one validated using its MERCA counterpart, returning the picture of countries focusing on a smaller set of products. For example, according to the export basket validated by $\mu-$MERCA$_c$, Tunisia is able to afford the production of items across the whole spectrum of goods, while such production is much more limited according to the export basket validated by $\mu-$BiWCM$_c$. An analogous consideration holds true for Jamaica.
	
	For what concerns $\alpha-$validated export baskets, MERCA-induced export baskets of small- and medium-fitness countries include more products of small complexity than the export baskets of high-fitness countries. BiWCM$_c$-induced export baskets complement this picture: while small- and medium-fitness countries still include products of small complexity, a strong signal concerning the presence of high-complexity products in the export basket of the high-fitness countries becomes now visible.
	
	\begin{table}[t!]
		%\begin{tabular}{c|cccc}
		%	RCA & 1 &  &  &  \\
		%	level 1 & 0.906 & 1  &  & \\
		%	level 2 & 0.434 & 0.616  & 1 &  \\
		%	level 3 & 0.434 & 0.616  & 1 & 1 \\
		%	\hline
		%	& RCA & level 1 & level 2 & level 3
		%\end{tabular}
		%\begin{tabular}{c|cccc}
		%        RCA &1&&&\\
		%        Level 1&0.974&1&&\\
		%        Level 2&0.932&0.951&1&\\
		%        Level 3&0.948&0.973&0.954&1\\
		%	\hline
		%        &RCA&Level 1&Level 2&Level 3\\
		%\end{tabular}
		FITNESS\\ 
		$\,$\newline
		\begin{tabular}{c|cccc}
			\hline\\
			$\textbf{M}^{\mu-\textnormal{MERCA}_c}$ = RCA &1&&&\\
			$\textbf{M}^{\alpha-\textnormal{MERCA}_c}$&{0.852}&1&&\\
			$\textbf{M}^{\mu-\textnormal{BiWCM}_c}$&0.974&{0.809}&1&\\
			$\textbf{M}^{\alpha-\textnormal{BiWCM}_c}$&0.948&{0.817}&0.973&1\\
			\hline
			&$\textbf{M}^{\mu-\textnormal{MERCA}_c}$&$\textbf{M}^{\alpha-\textnormal{MERCA}_c}$&$\textbf{M}^{\mu-\textnormal{BiWCM}_c}$&$\textbf{M}^{\alpha-\textnormal{BiWCM}_c}$\\
		\end{tabular}
		\\ 
		$\,$\newline
		$\,$\newline
		COMPLEXITY\\
		$\,$\newline
		\begin{tabular}{c|cccc}
			\hline\\
			$\textbf{M}^{\mu-\textnormal{MERCA}_c}$ = RCA &1&&&\\
			$\textbf{M}^{\alpha-\textnormal{MERCA}_c}$&{0.653}&1&&\\
			$\textbf{M}^{\mu-\textnormal{BiWCM}_c}$&0.733&{0.645}&1&\\
			$\textbf{M}^{\alpha-\textnormal{BiWCM}_c}$&0.478&{0.607}&0.584&1\\
			\hline
			&$\textbf{M}^{\mu-\textnormal{MERCA}_c}$&$\textbf{M}^{\alpha-\textnormal{MERCA}_c}$&$\textbf{M}^{\mu-\textnormal{BiWCM}_c}$&$\textbf{M}^{\alpha-\textnormal{BiWCM}_c}$\\
		\end{tabular}
		\caption{Spearman correlation coefficients among the Fitnesses and Complexities computed using the various validation procedure for the year 2015 of our dataset. The correlation is nearly constant during the entire dataset.}
		\label{tab:correlations}
	\end{table}

	\subsection{Comparing Fitnesses and Complexities calculated on differently validated matrices}
	
	Generally speaking, the values of fitness and complexity~\cite{Tacchella2012,Cristelli2013} used to reorder the rows and the columns of the validated matrices are different for different matrices; nevertheless, similarities exist. For example, the Spearman correlation coefficient between the vectors of fitnesses calculated on different matrices is always larger than 0.8, i.e. the ranking induced by the fitnesses on different, validated matrices is nearly the same; moreover, the values of the correlation coefficients are basically constant for the entire dataset, indicating that the (small) differences between fitness-induced rankings are quite stable. Still, the distribution of fitnesses obtained by running stricter statistical validations becomes more peaked in correspondence of the most important exporters (e.g. China). On the other hand, the Spearman correlation coefficient between the vectors of complexities is much lower, lying between 0.48 and 0.73 - a piece of evidence confirming the noisy character of complexities. Table~\ref{tab:correlations} reports the correlations for 2015 as the other years are characterised by very similar values.
	
	\subsection{Filtering real data}
	Null-models are often used to assess the significance of a given structure or signal in the network. In Economic Complexity, RCA is used to `filter' the signal of comparative advantage by discounting the size of countries and products in the global market. In this section, we provide a visual example of this filtering process carried out by BiWCM$_c$ on the same matrix widely used in this framework.
	
\begin{figure}
		\centering
		\includegraphics[width = \textwidth]{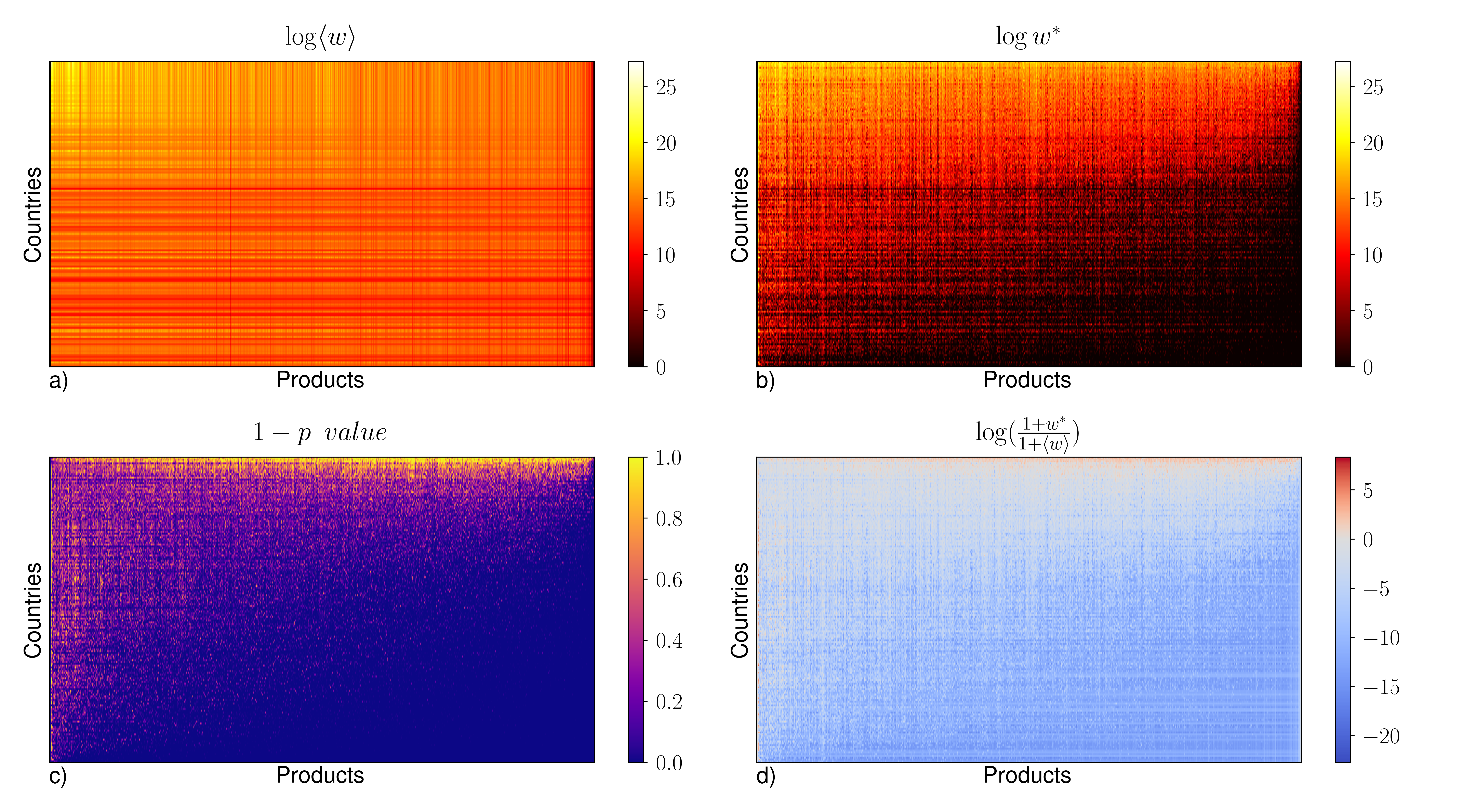}
		\caption{\textbf{Real Vs expected trade flows.} a) Expected value of trade flows given by Eq.(\ref{eq:likelihood_biwcm}). b) Real trade value. c) $1-\textnormal{p-value}_{\textnormal{BiWCM}_c}(w_{i\alpha})$ given by Eq.(\ref{eq:pval_biwcm}) as a measure of surprise of exceeding value of real data. d) Logarithm of the ratio $\frac{1+w^*}{1+\langle w \rangle_{\textnormal{BiWCM}_c}}$ where we added 1 to every entry to avoid infinities and the logarithm of zero issues. All matrices have the same ordering, given by the ranking of Fitness and Complexity calculated with the matrix in panel c).
		}
		\label{fig:res}
	\end{figure}

	In order to explore the properties of the BiWCM$_c$ in more detail, a comparison between the expected values  output by the model (panel a) and the empirical values (panel b) is displayed in Fig.~\ref{fig:res}.
	The most striking difference is the triangular and nested structure of the real data that completely vanishes in the expected values. As anticipated in the Methods section, this is not surprising: for each couple of non-isolated nodes, the expected weight is non-zero, see Eq.s(\ref{eq:likelihood_bimcm}) and (\ref{eq:likelihood_biwcm}). Such behaviour derives from the definition of the null-models considered, only discounting strength sequences that are continuous variables and not the node's degrees. Other entropy-based null-models available in the literature are able to discount both the information from topology (i.e. the node's degrees) and the weights (i.e. the strengths), see for instance~\cite{Mastrandrea2014}, although the topology is often inferred from the weights and is not available independently. 
	However, the scope of the present manuscript is to re-interpret Balassa's approach in a proper framework, keeping only the minimal null-model, while further developments will be scrutinized in future works.
	
	In panel c) shows the values of $1-\textnormal{p-value}_{\textnormal{BiWCM}_c}(w_{i\alpha})$ as a measure of signal's strength for trade: higher values mean an exceeding trade that is not explained simply by a random allocation of the strengths. 
	Countries and products are reordered on the basis of the Fitnesses and Complexities calculated on the 1-p-value matrix, for a visual impact of the picture. From Fig.~\ref{fig:res}, the most unexpected links of the network are those in the high-complexity regions. Such an effect is due to the nature of our model: BiWCM$_c$ distributes the strengths of the nodes among more links than those that actually exist in the real data, since $\langle w_{i\alpha}\rangle_{\textnormal{BiWCM}_c}\neq0$ for non-isolated nodes. High-complexity (low-ubiquity) products have very few links distributed among only high-fitness countries and their actual values are much higher than the predictions of BiWCM$_c$, see panel d). Remarkably, the order of Fitness and Complexity calculated on the matrix in panel c) and on the RCA matrix ($\mu$ validated MERCA) are highly correlated, with a Spearman correlation of 0.981.
	
	Summarising, so far we got two main insights about the model and system:
	BiWCM$_c$ can be used as a more refined, unbiased null-model to infer significant trades not explained simply by the strengths of nodes. Furthermore, as expected, it holds similar, but not equivalent, results to RCA in terms of shape of the matrix $1-\textnormal{p-value}$ and numerical values of Fitness and Complexity.

	%\end{comment}
	
	\section{Discussion and conclusions}
	
	The analysis of the bipartite representation of International Trade, placing countries on one layer and their exported products on the other one, provides particularly useful information to infer the industrial capabilities of countries~\cite{Hidalgo2009, Tacchella2012,Cristelli2013, Cristelli2015}. Typically, analyses are carried out on a binarised version of this data, that focuses on the export of products for which countries have a comparative advantage. According to Balassa's idea, a country $i$ that exports a selected product $\alpha$ `more than expected', under a model informed by just the total export of $i$ and the total export of $\alpha$, proves to have a comparative advantage for what concerns the export of $\alpha$~\cite{Balassa1965}. Balassa's implementation of the aforementioned idea prescribes comparing the empirical value of the export of $\alpha$ with a benchmark that is defined by the product of the total export of $i$, i.e. $s_i$, and the total export of $\alpha$, i.e. $\sigma_\alpha$.% (the so-called Revealed Comparative Advantage, or RCA, index).
	
	With the present manuscript, we have shown that following Balassa's rationale, it is possible to define an entropy-based benchmark preserving the same information advocated by Balassa: indeed, maximising Shannon entropy while constraining the total export of each country and the total export of each product~\cite{Cimini2018} leads to (both the discrete and the continuous variants of) the Bipartite Configuration Model for weighted networks (hereafter, \emph{BiWCM}). Remarkably, the discrete (or multilink) variant of the Bipartite Configuration Model and the one originally proposed by Balassa do not differ much in the so-called `sparse' regime (i.e. when the weights are `small').\\
	
	Two choices are, then, viable: implementing either the benchmark characterised by the simplest functional form (i.e. the original Balassa's one) or the one grounded in statistical theory, although mathematically more complicated (i.e. the one based upon the BiWCM). The difference between the two, however, is not merely formal: on the one hand, `dressing' Balassa's proposal with a probability distribution, following the same procedure as for the BiWCM, leads to a statistical model not (intended to) maximising the probability of observing the real matrix. In this sense, the `dressed' Balassa's proposal is a benchmark tailored on the observed system, still not `centred' on it.\\
	On the other hand, ``standard'' entropy-based null-models are unbiased in the sense explained in~\cite{Garlaschelli2008}: the condition returned by the likelihood maximisation (i.e. the maximisation of the probability of observing the real weighted bipartite network) agrees with the model definition.\\ 
	Such a difference has practical implications when implementing the different validation procedures considered in the present work.\\
	
	When calculating the fitnesses of countries, the error made by using plain RCA turns out to be limited as the rankings obtained using different methods are highly correlated; the situation is different when considering the complexities of products, whose rankings are less correlated. More specifically, the limits of the export baskets are neater in the matrix binarised by employing the BiWCM$_c$; such behaviour is captured by the nestedness of the biadjacency matrix: in fact, we observed that a BiWCM$_c$-induced binarisation returns a sNODF~\cite{Mariani2019} which is more than $10\%$ larger than the value measured on the matrix binarised according to the plain RCA.\\
	
	Our framework also allows us to implement proper statistical filtering by validating the p-values associated with the observations - in a sense, implementing a sort of `statistically significant comparative advantage' - instead of just retaining weights which are larger than expected. Countries' diversification is still evident; here, however, we also observe an excess of validated entries among the products exporting which a country is particularly efficient: specifically, while for `average' countries the statistically validated products are distributed across the entire export basket, for the fittest ones this effect is observed more frequently for the most complex products~\cite{Straka2017}.\\
	
	Summarising, we have provided a different interpretation of Balassa's recipe for comparative advantage and proposed rigorous statistical validation. Although embodying Balassa's general rationale, our approach is well grounded and clearly shows that using different procedures has implications for the interpretation of the data. Notably, the differences are minimal for the evaluation of the fitnesses of countries while they are more relevant for the evaluation of the complexities of products.
	
	\section*{References}
	\bibliographystyle{unsrt}
	\bibliography{biblio}

\begin{thebibliography}{10}

\bibitem{Hidalgo2009}
C\'esar~A. Hidalgo and Ricardo Hausmann.
\newblock The building blocks of economic complexity.
\newblock {\em Proceedings of the National Academy of Sciences of the United
  States of America}, 106:10570--10575, 6 2009.

\bibitem{Tacchella2012}
Andrea Tacchella, Matthieu Cristelli, Guido Caldarelli, Andrea Gabrielli, and
  Luciano Pietronero.
\newblock A new metrics for countries' fitness and products' complexity.
\newblock {\em Scientific Reports}, 2, 10 2012.

\bibitem{Cristelli2013}
Matthieu Cristelli, Andrea Gabrielli, Andrea Tacchella, Guido Caldarelli, and
  Luciano Pietronero.
\newblock Measuring the intangibles: A metrics for the economic complexity of
  countries and products.
\newblock {\em PLoS ONE}, 8, 2013.

\bibitem{Cristelli2015}
Matthieu Cristelli, Andrea Tacchella, and Luciano Pietronero.
\newblock The heterogeneous dynamics of economic complexity.
\newblock {\em PLoS ONE}, 10, 2 2015.

\bibitem{Balassa1965}
Bela Balassa.
\newblock Trade liberalisation and `revealed' comparative advantage.
\newblock {\em The Manchester School}, 33(2):99--123, 05 1965.

\bibitem{Balassa1967}
Bela Balassa.
\newblock Trade creation and trade diversion in the european common market.
\newblock {\em The Economic Journal}, 77(305):1--21, 03 1967.

\bibitem{Tacchella2018}
A.~Tacchella, D.~Mazzilli, and L.~Pietronero.
\newblock A dynamical systems approach to gross domestic product forecasting.
\newblock {\em Nature Physics}, 14:861--865, 8 2018.

\bibitem{Kunimoto1977}
Kazutaka Kunimoto.
\newblock Typology of trade intensity indices.
\newblock {\em Hitotsubashi Journal of Economics}, 17(2):15--32, 1977.

\bibitem{Vollrath1991}
Thomas~L. Vollrath.
\newblock A theoretical evaluation of alternative trade intensity measures of
  revealed comparative advantage.
\newblock {\em Weltwirtschaftliches Archiv}, 127(2):265--280, 1991.

\bibitem{krantz2018}
Ruben Krantz, Valerio Gemmetto, and Diego Garlaschelli.
\newblock Maximum-entropy tools for economic fitness and complexity.
\newblock {\em Entropy}, 20(10), 2018.

\bibitem{sbardella2018green}
Angelica Sbardella, Fran{\c{c}}ois Perruchas, Lorenzo Napolitano, Nicol{\`o}
  Barbieri, and Davide Consoli.
\newblock Green technology fitness.
\newblock {\em Entropy}, 20(10):776, 2018.

\bibitem{de2022trickle}
Francesco de~Cunzo, Alberto Petri, Andrea Zaccaria, and Angelica Sbardella.
\newblock The trickle down from environmental innovation to productive
  complexity.
\newblock {\em Scientific Reports}, 12(1):22141, 2022.

\bibitem{sbardella2017economic}
Angelica Sbardella, Emanuele Pugliese, and Luciano Pietronero.
\newblock Economic development and wage inequality: A complex system analysis.
\newblock {\em PloS one}, 12(9):e0182774, 2017.

\bibitem{patelli2023geography}
Aurelio Patelli, Lorenzo Napolitano, Giulio Cimini, and Andrea Gabrielli.
\newblock Geography of science: Competitiveness and inequality.
\newblock {\em Journal of Informetrics}, 17(1):101357, 2023.

\bibitem{pugliese2019unfolding}
Emanuele Pugliese, Giulio Cimini, Aurelio Patelli, Andrea Zaccaria, Luciano
  Pietronero, and Andrea Gabrielli.
\newblock Unfolding the innovation system for the development of countries:
  coevolution of science, technology and production.
\newblock {\em Scientific reports}, 9(1):16440, 2019.

\bibitem{digangi2018}
Domenico~Di Gangi, Fabrizio Lillo, and Davide Pirino.
\newblock Assessing systemic risk due to fire sales spillover through maximum
  entropy network reconstruction.
\newblock {\em Journal of Economic Dynamics and Control}, 94:117--141, 9 2018.

\bibitem{Garlaschelli2008}
Diego Garlaschelli and Maria~I. Loffredo.
\newblock Maximum likelihood: Extracting unbiased information from complex
  networks.
\newblock {\em Physical Review E - Statistical, Nonlinear, and Soft Matter
  Physics}, 78:1--5, 2008.

\bibitem{Cimini2018}
Giulio Cimini, Tiziano Squartini, Fabio Saracco, Diego Garlaschelli, Andrea
  Gabrielli, and Guido Caldarelli.
\newblock The statistical physics of real-world networks.
\newblock {\em Nature Reviews Physics}, 1:58--71, 1 2018.

\bibitem{Park2004}
Juyong Park and M.~E~J Newman.
\newblock Statistical mechanics of networks.
\newblock {\em Physical Review E - Statistical, Nonlinear, and Soft Matter
  Physics}, 70, 2004.

\bibitem{Parisi2020}
Federica Parisi, Tiziano Squartini, and Diego Garlaschelli.
\newblock A faster horse on a safer trail: generalized inference for the
  efficient reconstruction of weighted networks.
\newblock {\em New Journal of Physics}, 22:053053, 5 2020.

\bibitem{Mastrandrea2014}
Rossana Mastrandrea, Tiziano Squartini, Giorgio Fagiolo, and Diego
  Garlaschelli.
\newblock Enhanced reconstruction of weighted networks from strengths and
  degrees.
\newblock {\em New Journal of Physics}, 16, 7 2014.

\bibitem{Benjamini1995}
Yoav Benjamini and Yosef Hochberg.
\newblock Controlling the false discovery rate: a practical and powerful
  approach to multiple testing.
\newblock {\em Journal of the Royal Statistical Society B}, 57:289--300, 1995.

\bibitem{Saracco2017}
Fabio Saracco, Mika~J. Straka, Riccardo {Di Clemente}, Andrea Gabrielli, Guido
  Caldarelli, and Tiziano Squartini.
\newblock {Inferring monopartite projections of bipartite networks: An
  entropy-based approach}.
\newblock {\em New J. Phys.}, 2017.

\bibitem{Mariani2019}
Manuel~Sebastian Mariani, Zhuo-Ming Ren, Jordi Bascompte, and Claudio~Juan
  Tessone.
\newblock Nestedness in complex networks: Observation, emergence, and
  implications.
\newblock {\em Physics Reports}, 813:1--90, 2019.
\newblock Nestedness in complex networks: Observation, emergence, and
  implications.

\bibitem{AlmeidaNeto2008}
M\'ario Almeida-Neto, Paulo Guimar\~es, Paulo~R. Guimar\~aes Jr, Rafael~D.
  Loyola, and Werner Ulrich.
\newblock A consistent metric for nestedness analysis in ecological systems:
  reconciling concept and measurement.
\newblock {\em Oikos}, 117(8):1227--1239, 2008.

\bibitem{Straka2017}
M.J. Straka, G.~Caldarelli, and F.~Saracco.
\newblock Grand canonical validation of the bipartite international trade
  network.
\newblock {\em Physical Review E}, 96, 2017.

\bibitem{Dianati2016}
Navid Dianati.
\newblock A maximum entropy approach to separating noise from signal in bimodal
  affiliation networks.
\newblock {\em arXiv preprint arXiv:1607.01735}, 7 2016.

\bibitem{Vallarano2021}
Nicol\`o Vallarano, Matteo Bruno, Emiliano Marchese, Giuseppe Trapani, Fabio
  Saracco, Giulio Cimini, Mario Zanon, and Tiziano Squartini.
\newblock Fast and scalable likelihood maximization for exponential random
  graph models with local constraints.
\newblock {\em Scientific Reports}, 11, 12 2021.

\bibitem{Pugliese2016}
Emanuele Pugliese, Andrea Zaccaria, and Luciano Pietronero.
\newblock {On the convergence of the Fitness-Complexity algorithm}.
\newblock {\em The European Physical Journal Special Topics},
  225(10):1893--1911, 2016.

\bibitem{servedio2018new}
Vito~DP Servedio, Paolo Butt{\`a}, Dario Mazzilli, Andrea Tacchella, and
  Luciano Pietronero.
\newblock A new and stable estimation method of country economic fitness and
  product complexity.
\newblock {\em Entropy}, 20(10):783, 2018.

\bibitem{mazzilli2022fitness}
Dario Mazzilli, Manuel~Sebastian Mariani, Flaviano Morone, and Aurelio Patelli.
\newblock Fitness in the light of sinkhorn.
\newblock {\em arXiv preprint arXiv:2212.12356}, 2022.

\end{thebibliography}
	
	\newpage
	
	\appendix
	
	\section{Numerical methods to get Lagrange multipliers of the BiWCM$_d$}\label{app:BiMCM}
	
	As stated in the main text, to find the numerical values of the Lagrangian multipliers $\theta_i$ and $\eta_\alpha$ of the BiWCM$_d$, we have to maximize the log-likelihood of the observed strength sequence. %, and in order to solve this problem we implement Newton's method, following the same recipe of the binary bipartite model BiCM introduced by Vallarano et al.\cite{Vallarano2021}. 
	The log-likelihood reads:
	
	\begin{equation} \label{eq:loglikelihood_bimcm_basic}
		\mathcal{L} = \ln (P(G^*)) = \sum_{i, \alpha} \ln (1 - e^{-\theta_i-\eta_\alpha}) -\sum_i s_i^*\theta_i - \sum_\alpha \sigma_\alpha^* \eta_\alpha
	\end{equation}
	
	To maximize the likelihood in (\ref{eq:loglikelihood_bimcm_basic}), we compute its gradient and find its roots. Its gradient reads:
	
	\begin{equation}
		\frac{\partial \ln \mathcal{L}}{\partial \theta_i} = -s_i^* + \sum_\alpha \frac{e^{-\theta_i-\eta_\alpha}}{1-e^{-\theta_i-\eta_\alpha}} , \:\:\:
		\frac{\partial \ln \mathcal{L}}{\partial \eta_\alpha} = -\sigma_\alpha^* + \sum_i \frac{e^{-\theta_i-\eta_\alpha}}{1-e^{-\theta_i-\eta_\alpha}}.
	\end{equation}
	
	Indeed, finding the root of the gradient is equivalent to solve the constraint in Eq.~\ref{eq:likelihood_bimcm}. In the following, we will present the formulas for the Fixed-point and the Newton's methods to solve it. 
	
	\subsection*{Fixed-point}
	
	Numerically, the system of equations in (\ref{eq:likelihood_bimcm}) can be solved using the fixed-point method, i.e. using the iteration
	
	\begin{equation}\label{eq:fp_discrete}
		\left\{
		\begin{array}{c}
			\theta_i^{(n+1)}=-\ln\left[\frac{s_i^*}{\sum_\alpha\frac{e^{-\eta_\alpha^{(n)}}}{1-e^{-(\theta_i^{(n)}+\eta_\alpha^{(n)})}}}\right]=\theta_i^{(n)}+\ln\Bigg( \frac{\langle s_i^{(n)}\rangle}{s_i^*}\Bigg)=f_i(\vec{\theta}, \vec{\eta})\\ \\
			\eta_\alpha^{(n+1)}=-\ln\left[\frac{\sigma_\alpha^*}{\sum_i\frac{e^{-\theta_i^{(n)}}}{1-e^{-(\theta_i^{(n)}+\eta_\alpha^{(n)})}}}\right]=\eta_\alpha^{(n)}+\ln \Bigg(\frac{\langle \sigma_\alpha^{(n)}\rangle}{\sigma_\alpha^*}\Bigg)=f_\alpha(\vec{\theta}, \vec{\eta})
		\end{array}
		\right.
	\end{equation}
	until convergence, as proposed in Ref.s~\cite{Dianati2016,Vallarano2021}.\\
	
	The Jacobian associated to the map defined by the Eq.s~\ref{eq:fp_discrete} reads:
	\newcommand{\var}[1]{{\textnormal{Var}[#1]}}
	\newcommand{\cov}[1]{{\textnormal{Cov}[#1]}}
	
	\begin{equation}
		\left\{
		\begin{array}{l}
			\frac{df_i}{d\theta_j}=\delta_{ij}\Bigg(1-\frac{\sum_\alpha \var{w_{i\alpha}^{(n)}}}{\langle s_i^{(n)}\rangle}\Bigg)=\delta_{ij}\Bigg(1-\frac{\var{s_i^{(n)}}}{\langle s_i^{(n)}\rangle}\Bigg)\\ \\
			\frac{df_i}{d\eta_\alpha}=-\frac{\var{w_{i\alpha}^{(n)}}}{\langle s_i^{(n)}\rangle}=-\frac{\cov{s_i^{(n)},\sigma_\alpha^{(n)}}}{\langle s_i^{(n)}\rangle}\\ \\
			\frac{df_\alpha}{d\theta_i}=-\frac{\var{w_{i\alpha}^{(n)}}}{\langle \sigma_{\alpha}^{(n)}\rangle}=-\frac{\cov{s_i^{(n)},\sigma_\alpha^{(n)}}}{\langle \sigma_\alpha^{(n)}\rangle}\\ \\
			\frac{df_\alpha}{d\eta_\beta}=\delta_{\alpha\beta}\Bigg(1-\frac{\sum_i\var{w_{i\alpha}^{(n)}}}{\langle \sigma_\alpha^{(n)}\rangle}\Bigg)=\delta_{\alpha\beta}\Bigg(1-\frac{\var{\sigma_\alpha^{(n)}}}{\langle\sigma_\alpha^{(n)}\rangle}\Bigg)
		\end{array}
		\right.
	\end{equation}
	and $\var{w_{i\alpha}^{(n)}}=\frac{e^{-(\theta_i+\eta_\alpha)}}{[1-e^{-(\theta_i+\eta_\alpha)}]^2}$.\\
	
	\subsection*{Newton's method}
	
	To find the roots we employ Newton's method, so we compute the Hessian of the original log-likelihood, which reads
	
	\begin{equation}
		\left\{
		\begin{array}{c}
			\frac{\partial^2 \ln \mathcal{L}}{\partial \theta_i\partial\theta_j}=-\delta_{ij}\sum_\alpha\frac{e^{-(\theta_i+\eta_\alpha)}}{\big[1-e^{-(\theta_i+\eta_\alpha)}\big]^2} = -\delta_{ij}\sum_\alpha \var{w_{i\alpha}^{(n)}}= -\delta_{ij}\var{s_i^{(n)}}\\ \\
			\frac{\partial^2 \ln \mathcal{L}}{\partial \eta_\alpha\partial \eta_\beta} =-\delta_{\alpha\beta}\sum_i\frac{e^{-(\theta_i+\eta_\alpha)}}{\big[1-e^{-(\theta_i+\eta_\alpha)}\big]^2}= -\delta_{\alpha\beta}\sum_i \var{w_{i\alpha}^{(n)}}=-\delta_{\alpha\beta}\var{\sigma_\alpha^{(n)}}\\ \\
			\frac{\partial^2 \ln \mathcal{L}}{\partial \theta_i\partial\eta_\alpha} = \frac{\partial^2 \ln \mathcal{L}}{\partial \eta_\alpha\partial\theta_i}= -\frac{e^{-(\theta_i+\eta_\alpha)}}{\big[1-e^{-(\theta_i+\eta_\alpha)}\big]^2}=-\var{w_{i\alpha}^{(n)}}=-\cov{s_i^{(n)},\sigma_\alpha^{(n)}} 
		\end{array}
		\right.
	\end{equation}
	
	Newton's method is iterative and consists in employing the Hessian matrix for finding a direction for an update on the objective variable, in our case the Lagrangian multipliers. Our step then reads
	
	\begin{equation}\label{eq:gensol}
		\Delta(\vec{\theta}, \vec{\eta})^{(n)}=-{\mathbf{H}^{(n)}}^{-1}\nabla\mathscr{L}(\vec{\theta},\vec{\eta}).
	\end{equation}
	
	However, for finding the optimal solution for the Lagrange multipliers, this resolution method can be slow due to the high computational cost of inverting the Hessian matrix Newton's method. 
	Thus, for solving this problem we actually employ a \textit{quasi-Newton} approach, simplifying the Hessian matrix and considering only the elements on the diagonal, which are dominant with respect to the off-diagonal elements. Again, this strategy is inspired by the work for the binary bipartite model \cite{Vallarano2021} and it is the default method implemented as solvers in the aforementioned Python packages.

	\section{Numerical methods to get Lagrange multipliers of the BiWCM$_c$}\label{app:BiWCM}
	
	As in the previous section, we  maximize the log-likelihood of the observed strength sequence to find the numerical values of the Lagrangian multipliers $\theta_i$ and $\eta_\alpha$ of BiWCM$_c$.
	The log-likelihood reads:
	
	\begin{equation} \label{eq:loglikelihood_biwcm_basic}
		\mathcal{L}= \ln (P(G^*)) = \sum_{i, \alpha} \ln (\theta_i+\eta_\alpha) -\sum_i s_i^*\theta_i - \sum_\alpha \sigma_\alpha^* \eta_\alpha
	\end{equation}
	
	To maximize the likelihood in (\ref{eq:loglikelihood_biwcm_basic}), we compute its gradient and find its roots. Its gradient reads:
	
	\begin{equation}
		\frac{\partial \ln \mathcal{L}}{\partial \theta_i} = -s_i^* + \sum_\alpha \frac{1}{\theta_i+\eta_\alpha} , \:\:\:
		\frac{\partial \ln \mathcal{L}}{\partial \eta_\alpha} = -\sigma_\alpha^* + \sum_i \frac{1}{\theta_i+\eta_\alpha}.
	\end{equation}
	
	Again, finding the root of the gradient above is equivalent to solve the constraint in Eq.~\ref{eq:likelihood_biwcm}. In the following, we will present the formulas for the Fixed-point and the Newton's methods to solve it. 
	
	\subsection*{Fixed-point}
	In this case, we solve the system of equations iteratively, as has been already done for binary bipartite graphs \cite{Vallarano2021}. We can impose
	\begin{equation}\label{eq:likelihood_cont}
		\left\{
		\begin{array}{c}
			\theta_i^{(n+1)}=\frac{1}{s_i^*}\sum_\alpha\frac{1}{1+\frac{\eta_\alpha^{(n)}}{\theta_i^{(n)}}}=\theta_i^{(n)}\cdot\frac{\langle s_i^{(n)}\rangle}{s_i^*}=f_i(\vec{\theta}, \vec{\eta})\\ \\
			\eta_\alpha^{(n+1)}=\frac{1}{\sigma_\alpha^*}\sum_i\frac{1}{1+\frac{\theta_i^{(n)}}{\eta_\alpha^{(n)}}}=\eta_\alpha^{(n)}\cdot\frac{\langle \sigma_\alpha^{(n)}\rangle}{\sigma_\alpha^*}=f_\alpha(\vec{\theta}, \vec{\eta}).  
		\end{array}
		\right.
	\end{equation}
	
	The Jacobian associated to the transformation above reads:
	\begin{equation}
		\left\{
		\begin{array}{l}
			\frac{df_i}{d\theta_j}=-\frac{\delta_{ij}}{s_i^*}\sum_\alpha\frac{\eta_\alpha^{(n)}}{\big[\theta_i^{(n)}+\eta_\alpha^{(n)}\big]^2}=-\frac{\delta_{ij}}{s_i^*}\sum_\alpha\eta_\alpha^{(n)}\var{w_{i\alpha}^{(n)}}\\ \\
			\frac{df_i}{d\eta_\alpha}=-\frac{1}{s_i^*}\frac{\theta_i^{(n)}}{\big[\theta_i^{(n)}+\eta_\alpha^{(n)}\big]^2}=-\frac{\theta_i^{(n)}}{s_i^*}\var{w_{i\alpha}^{(n)}}\\ \\
			\frac{df_\alpha}{d\theta_i}=-\frac{1}{\sigma_\alpha^*}\frac{\eta_\alpha^{(n)}}{\big[\theta_i^{(n)}+\eta_\alpha^{(n)}\big]^2}=-\frac{\eta_\alpha^{(n)}}{\sigma_\alpha^*}\var{w_{i\alpha}^{(n)}}\\ \\
			\frac{df_\alpha}{d\eta_\beta}=-\frac{\delta_{\alpha\beta}}{\sigma_\alpha^*}\sum_i\frac{\theta_i^{(n)}}{\big[\theta_i^{(n)}+\eta_\alpha^{(n)}\big]^2}=-\frac{\delta_{\alpha\beta}}{\sigma_\alpha^*}\sum_i\theta_i^{(n)}\var{w_{i\alpha}^{(n)}}
		\end{array}
		\right.,
	\end{equation}
	with the identification $\var{w_{i\alpha}^{(n)}}=\frac{1}{\big[\theta_i^{(n)}+\eta_\alpha^{(n)}\big]^2}$.
	
	\subsection*{Newton's method}
	As mentioned above, the system of equations in (\ref{eq:likelihood_biwcm}) represents the maximisation of the log-likelihood associated with the probability defined in Eq.~\ref{eq:p_biwcm}. The Hessian associated with the transformation reads
	
	\begin{equation}
		\left\{
		\begin{array}{c}
			\frac{\partial^2 \ln \mathcal{L}}{\partial \theta_i\partial\theta_j} = -\delta_{ij}\sum_\alpha \frac{1}{(\theta_i+\eta_\alpha)^2}=-\delta_{ij}\sum_\alpha\var{w_{i\alpha}^{(n)}}=-\delta_{ij}\var{s_i^{(n)}}\\ \\
			\frac{\partial^2 \ln \mathcal{L}}{\partial \eta_\alpha^2} = -\delta_{\alpha\beta}\sum_i \frac{1}{(\theta_i+\eta_\alpha)^2}=-\delta_{\alpha\beta}\sum_i\var{w_{i\alpha}^{(n)}}=-\delta_{\alpha\beta}\var{\sigma_\alpha^{(n)}}\\ \\
			\frac{\partial^2 \ln \mathcal{L}}{\partial \theta_i\partial\eta_\alpha}=\frac{\partial^2 \ln \mathcal{L}}{\partial\eta_\alpha\partial \theta_i}  =-\frac{1}{(\theta_i+\eta_\alpha)^2}=-\var{w_{i\alpha}^{(n)}}=-\cov{s_i^{(n)},\sigma_\alpha^{(n)}}
		\end{array}
		\right.
	\end{equation}
	the generic step of Newton's method reading
	
	\begin{equation}\label{eq:gensol}
		\Delta(\vec{\theta}, \vec{\eta})^{(n)}=-{\mathbf{H}^{(n)}}^{-1}\nabla\mathscr{L}(\vec{\theta},\vec{\eta}).
	\end{equation}
	
	However, its convergence can still be slow. Thus, in this case, the default method we implemented in our solvers is the fixed-point strategy, which turns out to be faster and more accurate in the case of the BiWCM$_c$.

	\section{The Fitness and Complexity algorithm}\label{app:fitness}
	The Fitness and Complexity algorithm (FC) was developed by~\cite{Tacchella2012} and it is based on the definition of the Fitness of countries ( $F_c$ ) and of the Complexity of product ( $Q_p$ ) through the formulas:
	\begin{numparts}	
		\begin{equation}
			\tilde{F}_c^{(n)}=\sum_p M_{cp} Q_{p}^{(n-1)}
		\end{equation}
		\begin{equation}
			\tilde{Q}_{p}^{(n)}=\frac{1}{\sum_c M_{cp}\frac{1}{F_c^{(n-1)}}}
		\end{equation}
		\label{eq:fit-com}
	\end{numparts}
	with a normalization step at the end of each iteration:
	\begin{numparts}    \label{eq:fit-com-normalization}
		\begin{equation}
			F_c^{(n)}=\frac{\tilde{F}_c^{(n)}}{\left\langle\tilde{F}^{(n)}\right\rangle_c}
		\end{equation}
		\begin{equation}
			Q_p^{(n)}=\frac{\tilde{Q}_p^{(n)}}{\left\langle\tilde{Q}^{(n)}\right\rangle_p}
			\label{eq:complexity}
		\end{equation}
	\end{numparts}
	The matrix $M_{cp}$ is the binary representation of the bipartite network linking exporter country $c$ to the exported product $p$.
	This iterative algorithm converges to a fixed point of the coupled equations under some rather general conditions~\cite{Pugliese2016} and can be found in a more stable version in~\cite{servedio2018new}.
	The standard definition of Fitness is a country's diversification weighted by the complexity of products it exports, while the complexity is mostly determined by the lowest-fitness country that exports that commodity.
	Recent work describes the connection between this algorithm and the Sinkhorn-Knopp algorithm used in matrix scaling~\cite{mazzilli2022fitness}.

\end{document}